\documentclass[default]{aastex701}
\NewPageAfterKeywords
\newcommand{\pokemondlc}{66}
\newcommand{\pokemonnum}{520}
\newcommand{\multiplenum}{118}
\newcommand{\companionnum}{143}
\newcommand{\singlenum}{402}
\newcommand{\doublenum}{95}
\newcommand{\triplenum}{22}
\newcommand{\quadruplenum}{0}
\newcommand{\quintuplenum}{1}
\newcommand{\projseppeak}{7.91 au ($\sigma_{log(a)}=1.1$, SE$_{log(a)}=0.10$)}
\newcommand{\multiplicityrate}{$22.7\pm1.8\%$} 
\newcommand{\companionrate}{$27.5\pm2.0\%$} 

\begin{document}

\title{The POKEMON Speckle Survey of Nearby M dwarfs. IV. Distance-Limited Catalog (POKEMON-DLC)}

\correspondingauthor{Catherine A. Clark}
\email{clarkc@ipac.caltech.edu}

\author[0000-0002-2361-5812]{Catherine A. Clark}
\affil{NASA Exoplanet Science Institute, IPAC, California Institute of Technology, Pasadena, CA 91125, USA}
\email{clarkc@ipac.caltech.edu}

\author[0000-0003-4408-0463]{Zafar Rustamkulov}
\affiliation{IPAC, California Institute of Technology, Pasadena, CA 91125, USA}
\email{zafar@caltech.edu}

\author[0000-0002-8552-158X]{Gerard T. van Belle}
\affil{Astronomy and Planetary Science, Northern Arizona University, NAU Box 6010, Flagstaff, AZ, 86011-6010}
\email{gvanbelle@gmail.com}

\author[0000-0002-0885-7215]{Mark E. Everett}
\affiliation{NSF National Optical-Infrared Astronomy Research Laboratory, 950 N. Cherry Ave., Tucson, AZ 85719, USA}
\email{mark.everett@noirlab.edu}

\author[0000-0001-7746-5795]{Colin Littlefield}
\affiliation{Bay Area Environmental Research Institute, Moffett Field, CA 94035, USA}
\email{littlefield@baeri.org}

\author[0009-0002-9833-0667]{Sarah J. Deveny}
\affiliation{Bay Area Environmental Research Institute, Moffett Field, CA 94035, USA}
\affiliation{NASA Ames Research Center, Moffett Field, CA 94035, USA}
\email{deveny@baeri.org}

\author[0000-0002-5741-3047]{David R. Ciardi}
\affil{NASA Exoplanet Science Institute, IPAC, California Institute of Technology, Pasadena, CA 91125 USA}
\email{ciardi@ipac.caltech.edu}

\author[0000-0002-5823-4630]{Kaspar von Braun}
\affil{Lowell Observatory, 1400 West Mars Hill Road, Flagstaff, AZ 86001, USA}
\affil{Earth \& Planets Laboratory, Carnegie Science, 5241 Broad Branch Road, NW, Washington, DC 20015-1305}
\email{kaspar@lowell.edu}

\begin{abstract}

The Solar Neighborhood is dominated by stars smaller, colder, and fainter than the Sun: the M dwarfs. If we are to understand the context in which the Sun formed and evolved, then we must investigate the system architectures of our low-mass neighbors. We have therefore carried out the Pervasive Overview of Kompanions of Every M Dwarf in Our Neighborhood (POKEMON) speckle survey of nearby M-dwarf primaries. We created the survey with the goal of observing a volume-limited (north of -$30^{\circ}$) sample of M-dwarf primaries through M9 out to 15 pc at diffraction-limited resolution. Pre-Gaia parallax measurements yielded a catalog of 454 nearby M-dwarf primaries. However, the precise astrometry from Gaia indicated that there are additional low-mass sources within 15 pc. Here we present the POKEMON-Distance Limited Catalog (POKEMON-DLC), a supplemental catalog that consists of speckle observations for the \pokemondlc{} additional M-dwarf primaries identified by Gaia, increasing the number of ultracool dwarf (later than M6.5) primaries in the POKEMON catalog by a factor of 1.6. In our observations we detect four likely bound companions. After carrying out a literature search for additional companions, we update the projected separation distribution and find a peak at \projseppeak{}. We also update the M-dwarf stellar multiplicity and companion rates, and find values of \multiplicityrate{} and \companionrate{}, respectively. These results emphasize the utility of Gaia for identifying low-mass, nearby sources, and we find that ensuing characterization of these sources by SPHEREx will continue to clarify the nature of the Solar Neighborhood.

\end{abstract}

\keywords{stars: binaries: visual --- stars: imaging --- stars: low-mass --- stars: statistics --- solar neighborhood}


\section{Introduction}

Low-mass stars dominate the Solar Neighborhood \citep{Henry2006AJ....132.2360H, Winters2015AJ....149....5W}. The molecular clouds of gas and dust that collapse to form protostars prefer to fragment into many small pieces, rather than a few large ones \citep[e.g.,][]{McKeeOstriker2007ARA&A..45..565M}. As such, of the $\sim400$ stellar objects known to be within 10 parsecs (pc), not a single one is of spectral type O or B; in contrast, there are nearly 300 M dwarfs in this same volume \citep{HenryJao2024ARA&A..62..593H}.

However, despite the omnipresence of nearby M dwarfs, they can be difficult to find. Their flux peaks at near-infrared (NIR) wavelengths \citep[e.g., Figure 1 of][]{Shields2016PhR...663....1S}, so many were not discovered until the advent of NIR detectors in the mid-twentieth century \citep[a thorough review of the development of this technology is provided in][]{Rogalski2012OERv...20..279R}.

All-sky NIR surveys such as the Two Micron All-Sky Survey \citep[2MASS;][]{Skrutskie2006AJ....131.1163S} and the Wide-field Infrared Survey Explorer \citep[WISE;][]{Wright2010AJ....140.1868W}, as well as space missions such as the Infrared Space Observatory \citep{Kessler1996A&A...315L..27K} and Spitzer \citep{Werner2004ApJS..154....1W}, have greatly expanded our knowledge of the contents of the Solar Neighborhood, and in particular our low-mass neighbors. In fact, these catalogs are still being used to discover new nearby candidates \citep{Brooks2024AJ....168..211B, Karpov2025A&A...695A.195K}, and astrometric measurements from Gaia \citep{Gaia2016A&A...595A...1G} have clarified which of these candidates lie within a couple dozen pc of the Sun \citep{Scholz2020A&A...637A..45S}.

Earlier this year, the SpectroPhotometer for the History of the universe, Epoch of Reionization, and ices Explorer (SPHEREx) mission \citep{Bock2025arXiv251102985B} launched. SPHEREx is expected to find and characterize even more nearby, low-mass objects by producing a NIR spectrum for every six arcsecond pixel on the sky.

Critically, these missions have led (and will lead) to the discovery of many new nearby ultracool dwarfs \citep[UCDs;][]{Kirkpatrick1997AJ....113.1421K}, the faintest objects in the optical that still fuse hydrogen, found at the bottom of the Main Sequence \citep{BaraffeChabrier1996ApJ...461L..51B, Dieterich2014AJ....147...94D}. UCDs are typically defined as M dwarfs with spectral types later than M6.5 \citep{Gillon2024NatAs...8..865G}. Though challenging to find and confirm, UCDs -- and in particular the rate at which they host stellar companions -- are critical to our understanding of the initial mass function, which describes star formation throughout the Milky Way.

Furthermore, all but two of the 42 known planets within 5 pc orbit M dwarfs (and the two non-M-dwarf hosts are K dwarfs)\footnote[1]{https://exoplanetarchive.ipac.caltech.edu}. Stellar multiplicity is known to affect multiple facets of planet formation \citep[e.g.,][]{Hirsch2021AJ....161..134H}, detection \citep[e.g.,][]{Lester2021AJ....162...75L}, and characterization \citep[e.g.,][]{Ciardi2015ApJ...805...16C}, as well as resultant planetary system architectures \citep[e.g.,][]{Clark2022AJ....163..232C}, so a knowledge of M-dwarf -- and in particular UCD -- multiplicity is critical if we are to understand the frequency and properties of planets in the Solar Neighborhood.

As such, the Pervasive Overview of Kompanions to Every M-dwarf in Our Neighborhood (POKEMON) survey has sought to determine the stellar multiplicity of every M-dwarf primary north of $-30^{\circ}$ within 15 pc using diffraction-limited speckle imaging, and includes multiplicity measurements for brighter M dwarfs at larger distances. The first paper in the series \citep{Clark2022AJ....164...33C} presented newly discovered stellar companions from the survey. The second paper in the series \citep{Clark2024AJ....167...56C} presented speckle observations for 1125 nearby M dwarfs, and evaluated the effect of stellar multiplicity on Gaia astrometry and metrics. The third paper in the series \citep{Clark2024AJ....167..174C} presented the stellar multiplicity rate of M dwarfs within 15 pc, as well as a gap in the projected separation distributions for planet-hosting and non-planet-hosting M dwarfs.

While the sample selection described in the previous POKEMON papers was intended to be as complete as possible, results from Gaia have revolutionized our understanding of the Solar Neighborhood. As such, this fourth paper in the series presents the POKEMON-Distance Limited Catalog (POKEMON-DLC), which is a supplemental catalog of speckle observations for the \pokemondlc{} M-dwarf primaries identified using Gaia color and parallax measurements.

In Section \ref{sec:methods}, we describe our sample selection, observational routine, and data reduction pipeline. In Section \ref{sec:results}, we present the detected companions, the companions known to the literature, an updated projected separation distribution, and updated stellar multiplicity and companion rates for M dwarfs within 15 pc. In Section \ref{sec:discussion}, we discuss Gaia as an identifier of UCDs and how data from the upcoming SPHEREx mission can be used to further our understanding of nearby UCDs. Finally, in Section \ref{sec:conclusions}, we summarize our findings and discuss future directions of the POKEMON survey.

\section{Methods} \label{sec:methods}

In this section we describe our sample selection, observational routine, and data reduction procedure.

\subsection{Original POKEMON Sample Selection}

The creation of the POKEMON catalog has been discussed in detail in the previous three works in the series \citep{Clark2022AJ....164...33C, Clark2024AJ....167...56C, Clark2024AJ....167..174C}. Briefly, the POKEMON catalog was designed to be volume-limited through M9 out to 15 pc, with additional brighter (earlier-type) M dwarfs at larger distances. The targets are all northern ($\delta > -30^{\circ}$), as we used speckle cameras in Arizona to observe the M dwarfs at diffraction-limited resolution.

The POKEMON targets were drawn from a number of catalogs and surveys on nearby and low-mass stars \citep{GlieseJahreiss1991adc..rept.....G, Henry2006AJ....132.2360H, Skrutskie2006AJ....131.1163S, DupuyLiu2012ApJS..201...19D, Deshpande2013AJ....146..156D, Kirkpatrick2014ApJ...783..122K, LuhmanSheppard2014ApJ...787..126L, Magnier2015IAUGA..2257922M, Waters2015IAUGA..2256019W}. However, the catalog was constructed prior to the launch of the Gaia mission. As such, the parallaxes (and thus the distances) of these sources were not measured in a uniform way.

\subsection{The Gaia-Augmented POKEMON-DLC}

An examination of Gaia Data Release 3 \citep[DR3;][]{Gaia2023AA...674A...1G} shows that there are 606 late-type dwarfs north of -$30^{\circ}$ and within 15 pc based on a $B_p - R_p > 1.80$ cut as suggested by \citet{PecautMamajek2013ApJS..208....9P}. Of these, we determined that 85 are secondaries and not primaries, 45 are brown dwarfs, 3 are white dwarfs, 2 are background objects, and 1 is an RR Lyrae variable. This means that 470 are M-dwarf primaries. Of these, 404 were already in the POKEMON catalog, meaning that according to Gaia, there were \pokemondlc{} late-type primaries within 15 pc missing.

As noted in \citet{Clark2024AJ....167...56C}, Gaia is frequently unable to provide an astrometric solution for faint targets or close multiples. As such, the original POKEMON catalog does include an additional 50 targets that were not identified by Gaia, but were identified by one of the other catalogs or surveys identified previously, bringing the total number of targets in the POKEMON catalog to \pokemonnum. This emphasizes that although Gaia is a powerful tool for identifying nearby sources, other techniques are needed for a complete census of nearby objects.

\begin{figure*}
    \centering
    \includegraphics[width=\textwidth]{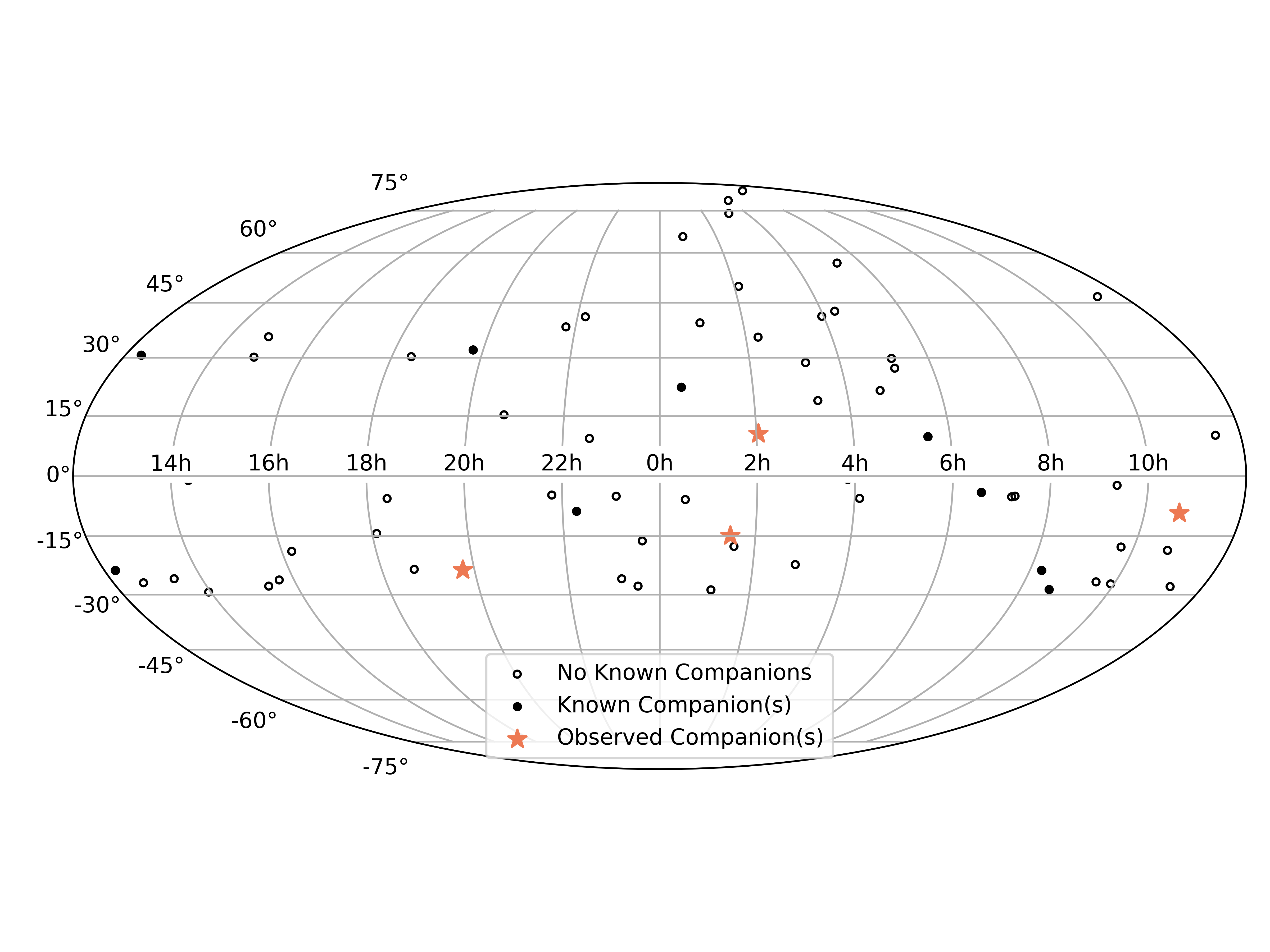}
    \caption{The sky locations of the \pokemondlc{} targets in the POKEMON-DLC supplemental catalog. Targets with no known companions are marked with open black circles, targets with known companions (but without companions detected by us) are marked with filled black circles, and targets with likely bound companions detected by us are marked with larger, orange stars.}
    \label{fig:aitoff}
\end{figure*}

In this work, we observed the remaining \pokemondlc{} M-dwarf primaries that were identified by Gaia as missing from the original POKEMON catalog (Figure \ref{fig:aitoff}). Table \ref{table:targets} includes the 2MASS ID or name, Gaia DR3 ID, $G$ magnitude, Gaia-parallax-derived distance, re-normalized unit weight error (RUWE) value, and the ipd\_frac\_multi\_peak (IPDFMP) value for each of these M dwarfs. We also note whether the target has any detected or known companion(s). Detected companions are discussed in Section \ref{subsec:detected} and Table \ref{table:detected}, and known companions are discussed in Section \ref{subsec:known} and Table \ref{table:known}.

\begin{deluxetable}{cccccccc}
\tablecaption{POKEMON-DLC targets
\label{table:targets}}
\tablehead{\colhead{2MASS ID or Name} & \colhead{Gaia DR3 ID} & \colhead{$G$ Magnitude} & \colhead{Distance} & \colhead{RUWE} & \colhead{IPDFMP} & \colhead{Companion(s) detected?} & \colhead{Known companion(s)?}\\ 
\colhead{} & \colhead{} & \colhead{(mag)} & \colhead{(pc)} & \colhead{} & \colhead{} & \colhead{} & \colhead{}}
\startdata
J00275592+2219328 & 2799992744809482112 & 15.0 & 14.1 & 6.4 & 1 & N & Y \\
J00313539-0552115 & 2526862202360516224 & 11.6 & 14.1 & 1.3 & 0 & N & N \\
J00492565+6518038 & 524551232110220800 & 13.2 & 14.3 & 1.2 & 0 & N & N \\
J00580115+3919111 & 373838214051410816 & 12.6 & 14.6 & 1.4 & 0 & N & N \\
J01081826-2848207 & 5032989500911228928 & 12.0 & 12.8 & 1.6 & 0 & N & N \\
J01283952-1458042 & 2452167910719793664 & 12.2 & 13.4 & 7.6 & 88 & Y & Y \\
J01335800-1738235 & 2450599697900838912 & 11.7 & 14.7 & 1.8 & 0 & N & N \\
PM J02024+1034 A & 2571309955621962240 & 11.9 & 14.4 & 3.5 & 38 & Y & Y \\
J02070382+4938441 & 358269546017619840 & 11.1 & 14.3 & 1.4 & 1 & N & N \\
J02170993+3526330 & 327944328126649856 & 13.7 & 10.3 & 1.2 & 0 & N & N \\
\enddata
\tablecomments{Table \ref{table:targets} is published in its entirety in the machine-readable format. A portion is shown here for guidance regarding its form and content.}
\end{deluxetable}

\begin{figure*}
    \centering
    \includegraphics[width=0.49\textwidth]{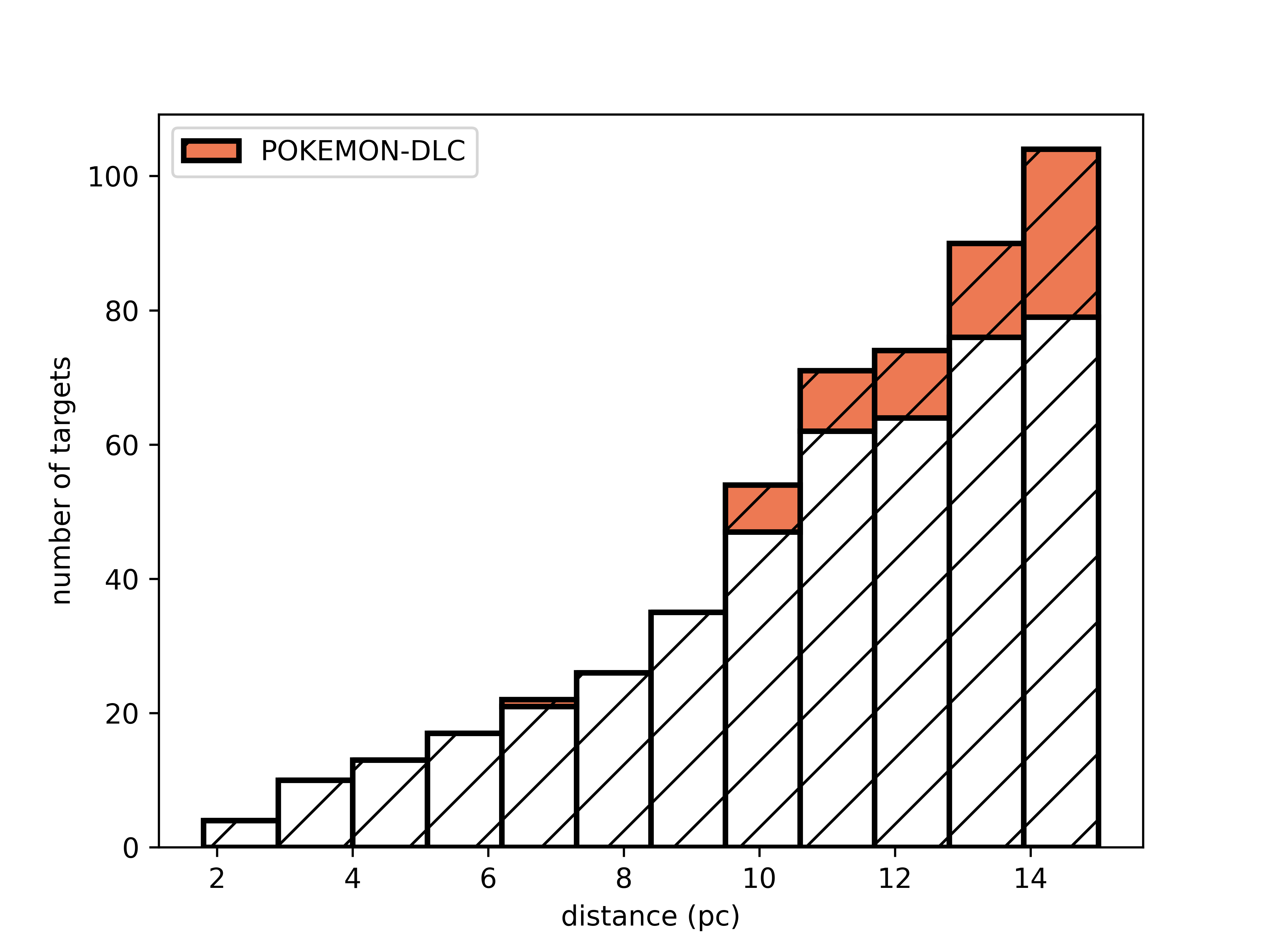}
     \includegraphics[width=0.49\textwidth]{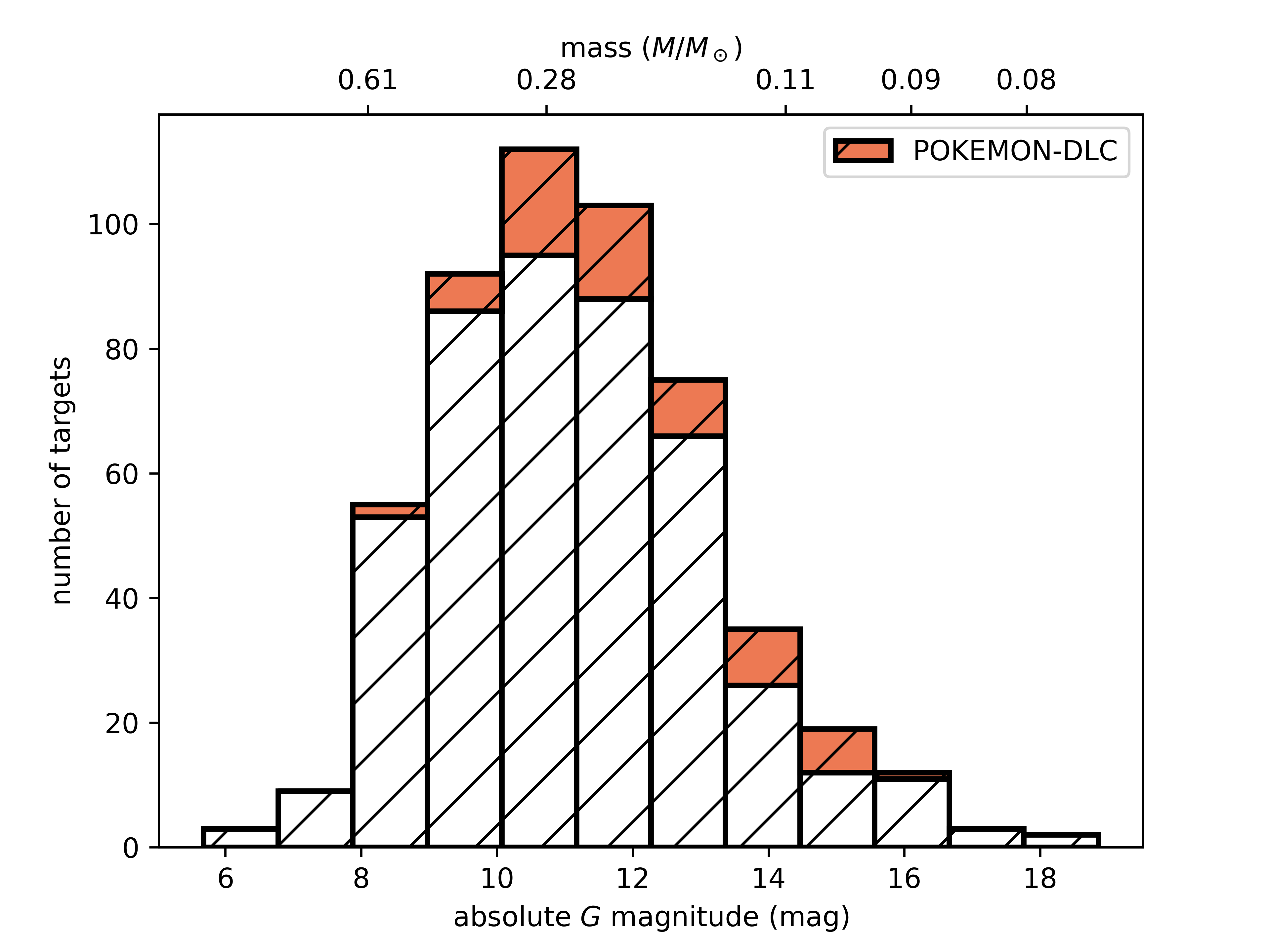}
    \caption{Distance (left) and absolute $G$ magnitude (right) distributions for the targets in the volume-limited, 15-pc POKEMON catalog. The POKEMON-DLC targets from this work are shown in orange. We use the mass-magnitude relation from \citet{HenryJao2024ARA&A..62..593H} to estimate the masses of these targets.}
    \label{fig:histograms}
\end{figure*}

In Figure \ref{fig:histograms} we show histograms of the distances and absolute $G$ magnitudes for the targets in the volume-limited, 15-pc POKEMON catalog. The POKEMON-DLC targets from this work are shown in orange. We use the absolute $G$ magnitudes and the mass-magnitude relation from \citet{HenryJao2024ARA&A..62..593H} to estimate masses for the POKEMON-DLC targets.

\subsection{Observational Routine and Data Reduction}

We imaged the \pokemondlc{} M-dwarf primaries in POKEMON-DLC using the NN-EXPLORE Exoplanet Stellar Speckle Imager \citep[NESSI;][]{Scott2018PASP..130e4502S} on the 3.5-meter WIYN telescope\footnote[2]{The WIYN Observatory is a joint facility of the NSF's National Optical-Infrared Astronomy Research Laboratory, Indiana University, the University of Wisconsin-Madison, Pennsylvania State University, the University of Missouri, the University of California-Irvine, and Purdue University.} at Kitt Peak National Observatory outside Tucson, Arizona. We used this instrument for consistency with the previous POKEMON observations. However, two targets (2MASS J15402966-2613422 and 2MASS J23312174-2749500) are both faint and near our southern declination cut, making them difficult to observe using NESSI. For these, we used the Zorro speckle camera \citep{Scott2021FrASS...8..138S} on the 8.1-meter Gemini South telescope. Zorro is nearly identical to NESSI, but is able to observe fainter objects due to the larger aperture.

For a detailed description of the speckle observations, we direct the reader to \citet{Clark2024AJ....167...56C}. Briefly, the collimated beam of light from the telescope is split at $\sim700$ nm using a dichroic filter. Each of the two channels is then directed through a narrowband filter and onto an electron-multiplying charge-coupled device. NESSI and Zorro use filters centered at 562 and 832 nm, with filter widths of 44 and 40 nm, respectively. Narrower bandpass filters provide more contrast in the speckle patterns; however, wider bandpass filters allow more photons to pass through, thereby improving the sensitivity of the instrument to faint companions. 40-50 nm filters were chosen to optimize this trade-off between contrast and sensitivity.

The raw data cubes from the instrument consist of 1,000 short exposures, on the order of tens of milli-seconds. This is the lifespan of an isoplanatic patch of atmosphere. Brighter objects require only one to three data cubes to be observed, while fainter objects require up to nine data cubes. Standard observing also includes observations of bright, unresolved, single stars from the Bright Star Catalog \citep{HoffleitJaschek1982bsc..book.....H} near in time and on-sky to the science target of interest. These observations are used to obtain a measurement of the Point Source Function as generated by the optical system and atmosphere at the time and location of the science target. These measurements are then used to reconstruct the image around each target.

The pixel scale and orientation are empirically confirmed using binaries with extremely well-known orbits \citep[those listed as Grade 1 in the Sixth Orbit Catalog;][]{Hartkopf2001AJ....122.3472H}. Their ephemeris positions are computed based on the orbital elements, and their scale and orientation are derived from these results \citep[][and references therein]{Horch2021AJ....161..295H}.

The data were reduced using an updated version of the pipeline described in \citet{Horch2009AJ....137.5057H, Horch2011AJ....141...45H, Horch2011AJ....141..180H}. Astrometric and photometric uncertainties are described in Sections 3.2 and 3.3, respectively, of \citet{Clark2024AJ....167...56C}. All data products (reconstructed images and contrasts) will be made publicly available on the Exoplanet Follow-up Observing Program archive\footnote[3]{https://exofop.ipac.caltech.edu}.

\section{Results} \label{sec:results}

In this section, we report the stellar companions that were detected in our speckle images and those known to the literature. We also provide an updated projected separation distribution and stellar multiplicity and companion rates for M dwarfs within 15 pc.

\subsection{Detected Companions} \label{subsec:detected}

\begin{figure}
    \centering
    \includegraphics[width=\textwidth]{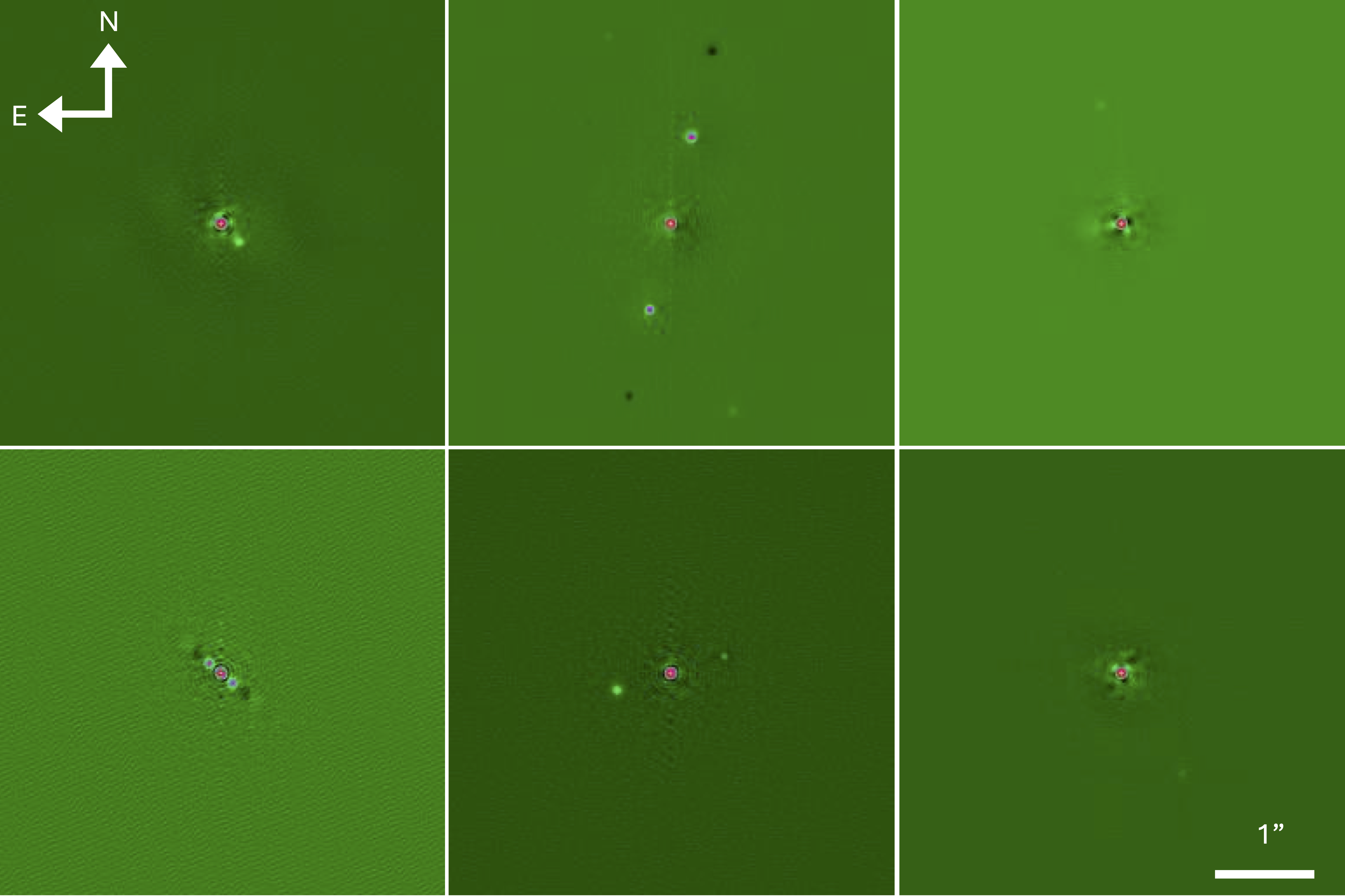}
    \caption{Starting from the top left corner and moving in a clockwise direction are the companions we detected to 2MASS J01283952-1458042, PM J02024+1034 A, 2MASS J05103956+2946479, 2MASS J07171706-0501031, 2MASS J10430293-0912410, and 2MASS J19445376-2337591. These reconstructed images sometimes contain a third ``ghost peak'' due to a low signal-to-noise ratio (SNR) resulting in a lack of phase information. As noted in Section \ref{subsec:detected}, the companions to 2MASS J05103956+2946479 and 2MASS J07171706-0501031 are unbound.}
    \label{fig:detections}
\end{figure}

We detected six companions to the \pokemondlc{} M dwarfs observed as part of POKEMON-DLC (Figure \ref{fig:detections}). Table \ref{table:detected} lists the 2MASS ID or name, Washington Double Star \citep[WDS;][]{Mason2001AJ....122.3466M, Mason2023yCat....102026M} ID, date of observation, central wavelength ($\lambda$), bandwidth ($\Delta\lambda$), position angle ($\theta$), angular separation ($\rho$), and delta magnitude ($\Delta m$) of the observations. We note several considerations for Table \ref{table:detected} below:

\begin{itemize}
    \item 2MASS J01283952-1458042 has hints of an additional, inner companion. After subtracting the model fringes from the observed fringes, residual structure is visible. This structure may be a single fringe of a barely resolved component, or simply an artifact due to high airmass during the observation. The separation would be approximately 0.06” if real.
    \item 2MASS J05103956+2946479 exhibits linear, rather than orbital, motion based on previous observations \citep{Cortes-Contreras2017AA...597A..47C, Lamman2020AJ....159..139L}, and is therefore unbound.
    \item In 2025, 2MASS J07171706-0501031 moved close ($\sim1.2\arcsec$) to a background star (Gaia DR3 3059246454092369280) that has a similar magnitude difference ($G\sim4$) to the detected companion. As such, we determine that this companion is unbound.
    \item There is a $180^{\circ}$ ambiguity in the position angle of 2MASS J19445376-2337591 due to the limitations of the reconstructed images as described in Section 2.3 of \citet{Clark2024AJ....167...56C}.
    \item As shown in previous works \citep[e.g.,][]{Horch2015AJ....150..151H, Horch2020AJ....159..233H}, when the seeing value multiplied by the angular separation is larger than 0.6'' squared, there may be a systematic error in the photometry, and the observed magnitude difference should be considered as an upper limit; i.e., the companions may be brighter than the quoted delta magnitude value. This is denoted by a $<$ limit flag on the delta magnitude.
\end{itemize}

\begin{deluxetable}{cccccccc}[h]
\tablecaption{Properties of detected companions
\label{table:detected}}
\tablehead{\colhead{2MASS ID or name} & \colhead{WDS ID} & \colhead{Epoch} & \colhead{$\lambda$} & \colhead{$\Delta\lambda$} & \colhead{$\theta$} & \colhead{$\rho$} & \colhead{$\Delta m$} \\ 
\colhead{} & \colhead{} & \colhead{(YYYY-MM-DD)} & \colhead{(nm)} & \colhead{(nm)} &\colhead{($^{\circ}$)} & \colhead{($\arcsec$)} & \colhead{(mag)}}
\startdata
J01283952-1458042 & 01287-1458 & 2024-09-10 & 832 & 40 & 224.3 & 0.273 & 2.58 \\
PM J02024+1034 A & 02025+1035 & 2024-09-10 & 562 & 44 & 346.5 & 0.924 & $<0.89$ \\
 &  & 2024-09-10 & 832 & 40 & 346.7 & 0.929 & $<0.59$ \\
J05103956+2946479\tablenotemark{*} & 05107+2947 & 2025-01-20 & 562 & 44 & 9.7 & 1.257 & $<3.92$ \\
 &  & 2025-01-20 & 832 & 40 & 9.7 & 1.272 & $<4.93$ \\
J07171706-0501031\tablenotemark{*} &  & 2025-01-20 & 562 & 44 & 211.3 & 1.227 & $<2.49$ \\
 &  & 2025-01-20 & 832 & 40 & 211.2 & 1.243 & $<4.60$ \\
J10430293-0912410 & 10430-0913 & 2025-01-20 & 832 & 40 & 107.5 & 0.595 & 2.14 \\
J19445376-2337591 & 19449-2338 & 2024-09-10 & 832 & 40 & 50.4 & 0.160 & 0.23 \\
\enddata
\tablenotetext{*}{Exhibits linear, rather than orbital, motion}
\tablecomments{Astrometric and photometric uncertainties are described in Sections 3.2 and 3.3, respectively, of \citet{Clark2024AJ....167...56C}.}
\end{deluxetable}

\subsection{Known Companions} \label{subsec:known}

Most stellar companions to M dwarfs are within 50 au \citep[][Figure 4 of this work]{Winters2019AJ....157..216W}, and the diffraction-limited nature of speckle imaging is well-suited to detecting stellar companions at these separations around nearby stars. In fact, speckle imaging recovers nearly four fifths of the stellar companions to M dwarfs within 15 pc \citep[Figure 3 of][]{Clark2024AJ....167..174C}.

However, speckle imaging only probes one area of parameter space, and M dwarfs are known to have companions at thousands of au. As such, we carried out a literature search to ensure the inclusion of all known stellar companions to the \pokemondlc{} M dwarfs in POKEMON-DLC. We mainly used the WDS and the Simbad bibliography tool \citep{Wenger2000A&AS..143....9W}, though we also carried out our own literature search for completeness and to ensure that the detected companions are bound.

In Table \ref{table:known} we report the first observed properties of the known stellar companions to the \pokemondlc{} M dwarfs in POKEMON-DLC. These are all M-dwarf companions; we do not consider any substellar companions such as brown dwarfs or planets. We also do not consider M dwarfs that have more massive companions, such as higher-mass Main Sequence stars, evolved stars, or white dwarfs. We note that in some cases these properties are quite different from those obtained more recently due to orbital motion.

Table \ref{table:known} includes the 2MASS ID or name of the primary, component configuration for each row, year of the measurement, position angle ($\theta$), angular separation ($\rho$), projected separation ($r$), detection technique, reference for the detection technique, magnitude difference ($\Delta m$), filter of the magnitude difference, and reference for the magnitude difference. We also list uncertainties on position angle, angular separation, and magnitude difference when available. For 2MASS J05321467+0949150, a spectroscopic binary, the projected separation was taken from \citet{Shkolnik2010ApJ...716.1522S}. There are $n-1$ lines for each system, where $n$ is the number of components in the system. In total, we find that 13 of the \pokemondlc{} POKEMON-DLC targets host stellar companions. Of these, 11 are double and 2 are triple. We note that, based on the RUWE and IPDFMP values listed in Table \ref{table:targets}, all POKEMON-DLC targets with elevated Gaia metrics indicative of stellar multiplicity have known stellar companions, increasing our confidence that all stellar multiples have been identified.

\begin{deluxetable}{cccccccccccccc}
\tablecaption{Properties of known companions}
\label{table:known}
\tablehead{\colhead{ID} & \colhead{Component} & \colhead{Epoch} & \colhead{$\theta$} & \colhead{Error} & \colhead{$\rho$} & \colhead{Error} & \colhead{$r$} & \colhead{Technique} & \colhead{Reference} & \colhead{$\Delta m$} & \colhead{Error} & \colhead{Filter} & \colhead{Reference} \\ 
\colhead{} & \colhead{} & \colhead{(yr)} & \colhead{($^{\circ}$)} & \colhead{($^{\circ}$)} & \colhead{($\arcsec$)} & \colhead{($\arcsec$)} & \colhead{(au)} & \colhead{} & \colhead{} & \colhead{(mag)} & \colhead{(mag)} & \colhead{} & \colhead{}}
\startdata
J00275592+2219328 & A-B & 2004 & 13 & 2 & 0.125 & 0.01 & 1.77 & AO det & 3 & 0.26 & 0.05 & K & 3 \\
J01283952-1458042\tablenotemark{*} & A-B & 2018 & 339 &  & 0.4115 &  & 5.49 & spkdet & 13 & 2.6 &  & I & 13 \\
PM J02024+1034 A\tablenotemark{*} & A-B & 2019 & 183 & 0.8 & 0.93 & 0.01 & 13.36 & AO det & 10 & 0.62 & 0.07 & I & 10 \\
J05321467+0949150 & A-B & 2009 &  &  &  &  & $<0.23$ & SB2 & 11 &  &  &  &  \\
J06352986-0403185 & A-B & 2012 & 171 & 1.2 & 0.155 & 0.006 & 1.96 & CCDdet & 5 & 2.28 & 0.1 & I & 5 \\
J08151117-2344157 & A-B & 2018 &  &  & 1.5 &  & 14.85 & astdet & 4 &  &  &  &  \\
J08383373-2843261 & A-B & 2019 & 98 &  & 0.0478 &  & 0.65 & spkdet & 13 & 0 &  & I & 13 \\
J10430293-0912410\tablenotemark{*} & A-B & 2006 & 267 &  & 0.477 &  & 5.82 & spkdet & 8 & 1.7 &  & V & 8 \\
J12140866-2345172 & A-B & 2016 & 223 & 0.01 & 3.4673 & 0.00072 & 37.33 & spcdet & 2 & 8.283 &  & Gaia & 2 \\
J12212673+3038376 & A-B & 1963 & 102 &  & 4.5 &  & 53.92 & phodet & 7 & 0.1 &  & B & 7 \\
J19445376-2337591\tablenotemark{*} & A-B & 2004 & 355 & 0.01 & 0.813 & 0.005 & 12.06 & AO det & 9 & 0.08 & 0.01 & K & 9 \\
J19462386+3201021 & A-B & 1935 & 135 &  & 2.21 &  & 30.82 & micdet & 12 & 1 &  & V & 12 \\
J22171899-0848122 & A-BaBb & 1920 & 225 &  & 7 &  & 77.36 & phodet & 6 & 0.8 &  & B & 6 \\
 & Ba-Bb & 2001 & 306 &  & 0.978 &  & 10.81 & AO det & 1 & 1.18 &  & K & 1
\enddata
\tablenotetext{*}{Additional detection(s) presented in Section \ref{subsec:detected}}
\tablecomments{The codes for the techniques and instruments used to detect and resolve systems are: AO det—detection via adaptive optics; astdet—detection via astrometric perturbation, companion often not detected directly; CCDdet—detection via CCD or other two-dimensional electronic imaging; micdet-detection via a micrometry technique; phodet—detection via a photographic technique; SB—spectroscopic multiple; spcdet-detection via space-based technique; spkdet—detection via speckle interferometry.}
\tablerefs{(1) \citet{Beuzit2004AA...425..997B};
(2) \citet{El-Badry2021MNRAS.506.2269E};
(3) \citet{Forveille2005AA...435L...5F};
(4) \citet{Henry2018AJ....155..265H};
(5) \citet{Janson2014ApJ...789..102J};
(6) \citet{Luyten1941POMin...3....1L};
(7) \citet{Luyten1969PMMin..21....1L};
(8) \citet{Mason2018AJ....155..215M};
(9) \citet{Montagnier2006AA...460L..19M};
(10) \citet{Salama2022AJ....163..200S};
(11) \citet{Shkolnik2010ApJ...716.1522S};
(12) \citet{vandeKamp1936PASP...48..313V};
(13) \citet{Vrijmoet2022AJ....163..178V}.}
\end{deluxetable}

\subsection{Updated Projected Separation Distribution and Stellar Multiplicity and Companion Rates for M Dwarfs within 15 pc}

In Section 3 of \citet{Clark2024AJ....167..174C}, we found that 107 of the 455 POKEMON targets within 15 pc hosted stellar companions. However, the ``companion'' to 2MASS J09510964-1219478 was found to be a double cosmic ray hit on one of the 5000 40-millisecond exposures \citep{Clark2024AJ....168..228C, Clark2024AJ....168..229C}, and the companion to 2MASS J12362870+3512007 is a brown dwarf \citep{Winters2020AJ....159..290W}. Additionally, 2MASS J07464256+2000321 was removed from the catalog completely, as it is a brown dwarf \citep{Kirkpatrick2000AJ....120..447K}. As such, we update these values to 105 companions and 454 targets.

\begin{figure}
    \centering
    \includegraphics[width=\textwidth]{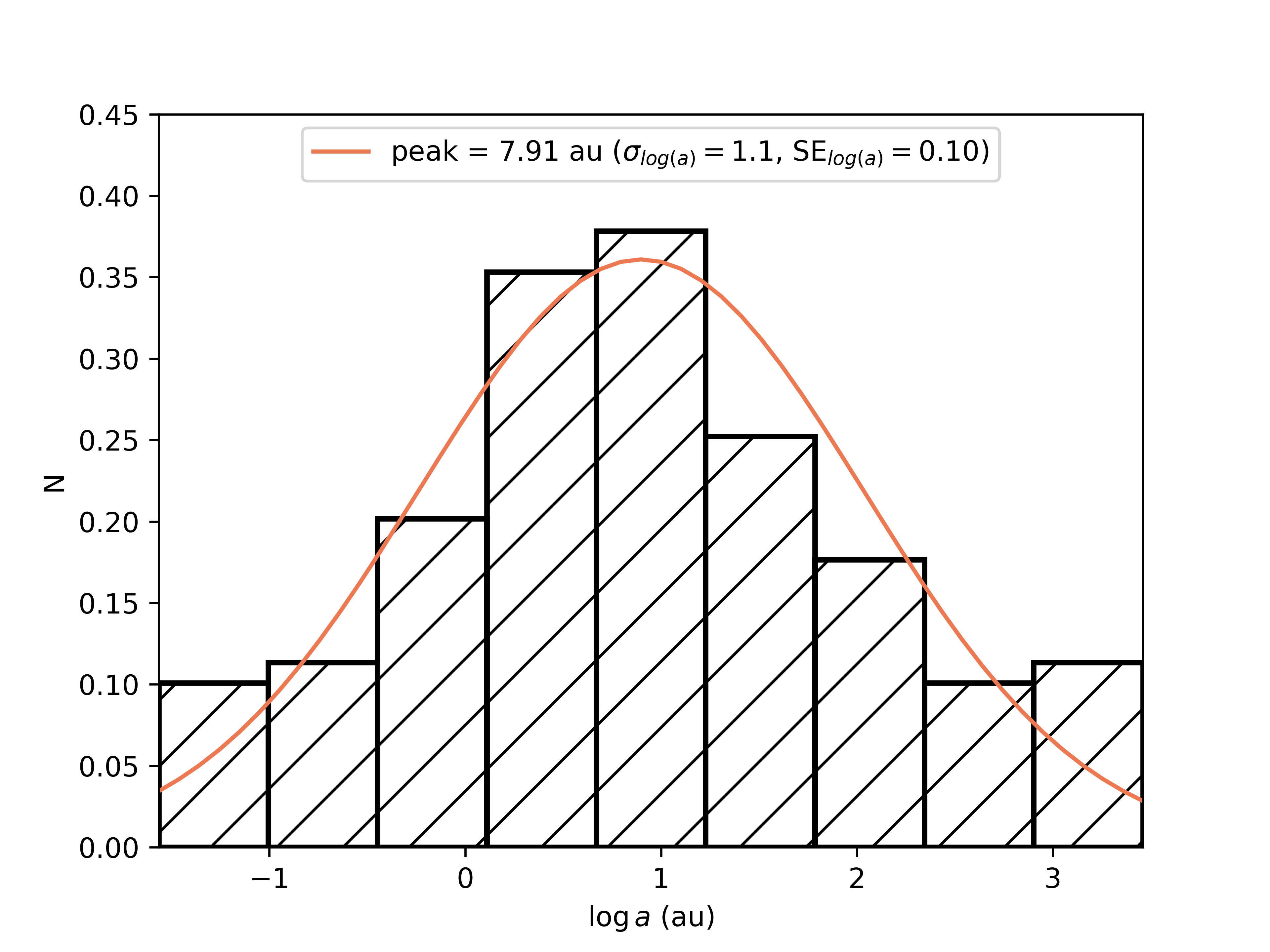}
    \caption{Projected separation distribution for the \companionnum{} stellar companions to the M dwarfs in the 15-pc POKEMON catalog. The peak of the distribution is at \projseppeak{}, which is consistent with other studies of M dwarf multiplicity \citep[i.e.,][]{DucheneKraus2013ARA&A..51..269D, Janson2014ApJ...789..102J, Winters2019AJ....157..216W}.}
    \label{fig:proj_sep_distribution}
\end{figure}

Combined with the 13 companions identified in Section \ref{subsec:known}, we find that there are \multiplenum{} stellar companions to the \pokemonnum{} M-dwarf primaries in the supplemented POKEMON catalog. Of these, \doublenum{} are double ($N_D$), \triplenum{} are triple ($N_T$), \quadruplenum{} are quadruple ($N_{Qd}$), and \quintuplenum{} is quintuple ($N_{Qn}$), for a total of \companionnum{} companions. As shown in Figure \ref{fig:proj_sep_distribution}, the companions have a projected separation distribution that is roughly Gaussian with a peak at \projseppeak. This distribution is in line with previous studies of M-dwarf multiplicity; \citet{DucheneKraus2013ARA&A..51..269D} found a peak at 5.3 au, and \citet{Janson2014ApJ...789..102J} found a peak at 6 au. \citet{Winters2019AJ....157..216W} found a peak at 20 au, but recognized that their search was not complete at small separations.

In addition to the projected separation distribution of these companions, we calculate updated stellar multiplicity and companion rates (MR and CR, respectively) using the following formulae from \citet{Winters2019AJ....157..216W}:

\begin{equation}
    MR = 100 * \frac{N_D + N_T + N_{Qd} + N_{Qn}}{N_S + N_D + N_T + N_{Qd} + N_{Qn}}
\end{equation}

\begin{equation}
    CR = 100 * \frac{N_D + 2N_T + 3N_{Qd} + 4N_{Qn}}{N_S + N_D + N_T + N_{Qd} + N_{Qn}}
\end{equation}

The multiplicity rate is simply defined as the percentage of systems that are multiple. In comparison, the companion rate is defined as the average number of companions per primary, so it takes into account whether the systems are binary, trinary, or of even higher orders.

We calculate the uncertainty on our rates using the analytic formula for the standard deviation of a binomial distribution

\begin{equation}
    \sigma_n = \sqrt{\frac{p (1 - p)}{n}}
\end{equation}

\noindent where $n$ is the number of experiments (the number of primaries) and $p$ is the probability of success (the number of multiples divided by the number of primaries).

We find an updated stellar multiplicity rate of \multiplicityrate{} and a stellar companion rate of \companionrate{} for M dwarfs within 15 pc. We also find that the ratio of singles:doubles:triples:higher-order systems is \singlenum:\doublenum:\triplenum:\quadruplenum:\quintuplenum, corresponding to 77.3:18.3:4.2:0.2\%. These findings are consistent with other recent studies of M-dwarf multiplicity. \citet{Jodar2013MNRAS.429..859J} surveyed a sample of dwarfs from late K to mid M using lucky imaging and found a rate of $20.3^{+6.9}_{-5.2}\%$. Soon after \citet{Janson2014ApJ...789..102J} surveyed late M-dwarfs ($>$M5) also using lucky imaging and found a rate of 21-27\%. \citet{Ward-Duong2015MNRAS.449.2618W} also surveyed the mass range from late K to mid M, but with AO, and found a rate of $23.5\pm3.2\%$.

In Appendix \ref{appendix}, we update relevant tables from \citet{Clark2024AJ....167...56C} and \citet{Clark2024AJ....167..174C} to include the POKEMON-DLC targets. Table \ref{table:all_targets} lists all \pokemonnum{} targets in the catalog and whether they have detected or known stellar companions. Detected companions are listed in Table \ref{table:all_detected}, and known companions are listed in Table \ref{table:all_known}. In addition to the newly added targets, Table \ref{table:all_detected} now includes one corrected WDS identifier, and Table 6 now includes corrections to several typographical errors in the projected separation values, which had been propagated into the projected separation distribution. We hope that this volume-limited census of nearby M-dwarf multiples will be of use to both the stellar astrophysics and exoplanet communities interested in the Solar Neighborhood.

\section{Discussion} \label{sec:discussion}

In this section we examine the use of Gaia for identifying late-type dwarfs within 15 pc. We also discuss expectations for how SPHEREx will our better our understanding of our low-mass neighbors.

\subsection{Gaia As an Identifier of Nearby, Ultracool Dwarfs}

According to a search of the full-sky, 20-pc census from \citet{Kirkpatrick2024ApJS..271...55K}, there are 19 primaries with $0.077<m<0.093 M_{\odot}$ \citep[corresponding to spectral types between M6.5 and L0;][]{PecautMamajek2013ApJS..208....9P} within 15 pc and north of $\delta=-30^{\circ}$. 12 of these were in the original 15-pc POKEMON catalog \citep{Clark2024AJ....167...56C}, and seven were observed as a part of this work, increasing the number of UCDs in the 15-pc POKEMON catalog by a factor of 1.6. As such, more than a third of these UCDs were not known to be within 15 pc prior to the launch of Gaia. These results demonstrate that the astrometry from Gaia is critical to our knowledge of nearby stars, including those at the end of the Main Sequence.

Furthermore, Figure 26 of \citet{Smart2021} shows that Gaia is 100\% complete for M dwarfs through M9 out to several dozen pc. As such, by using Gaia to create POKEMON-DLC, we can be confident that the POKEMON catalog now contains every M-dwarf primary through M9 out to 15 pc.

\begin{figure}
    \centering
    \includegraphics[width=0.9\linewidth]{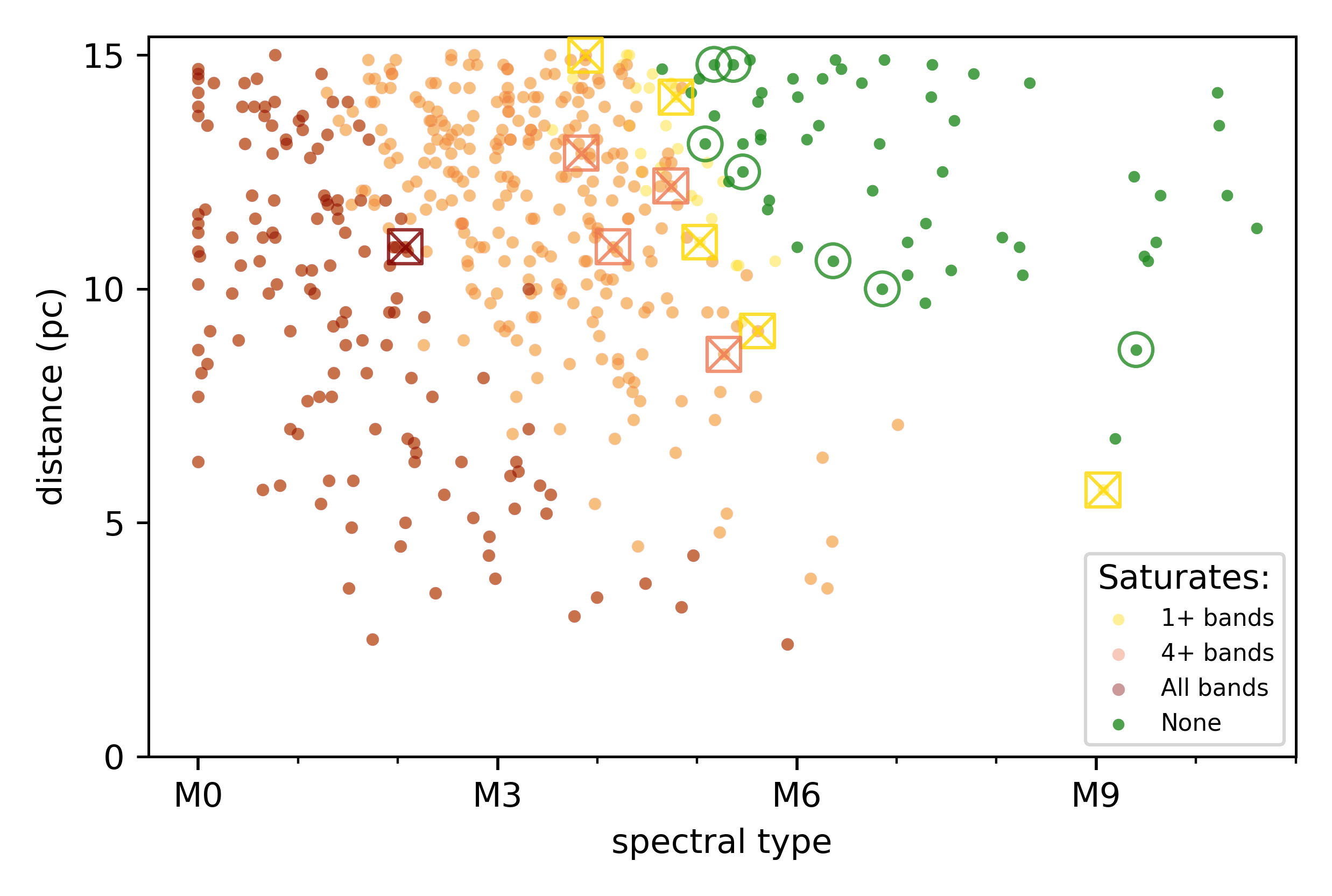}
    \caption{The POKEMON catalog as a function of spectral type and distance, colored by the level of saturation across the six SPHEREx bands. All stars earlier than spectral type M4.5 saturate at least one band. Points encapsulated in green circles were manually spot-checked using the IRSA SPHEREx Spectrophotometry Tool and showed few or no saturation flags in the brightest image pixels. Boxed points showed saturation flags and unreliable fluxes in one or more bands, corroborating the saturation estimates reflected in the point colors.}
    \label{fig:saturation}
\end{figure}

We note that none of the UCDs in either the original 15-pc POKEMON catalog nor POKEMON-DLC were found to have detected or known companions. This result supports the observation that stellar multiplicity decreases as a function of stellar mass \citep[e.g,][and references therein]{Offner2023ASPC..534..275O}.

\subsection{SPHEREx As a Characterizer of Nearby, Ultracool Dwarfs}

Although Gaia is a profoundly powerful mission, the extent to which it can be used to characterize low-mass stars is more limited. However, using Gaia in combination with other space-based missions can yield powerful results.

\begin{figure}
    \centering
    \includegraphics[width=\linewidth]{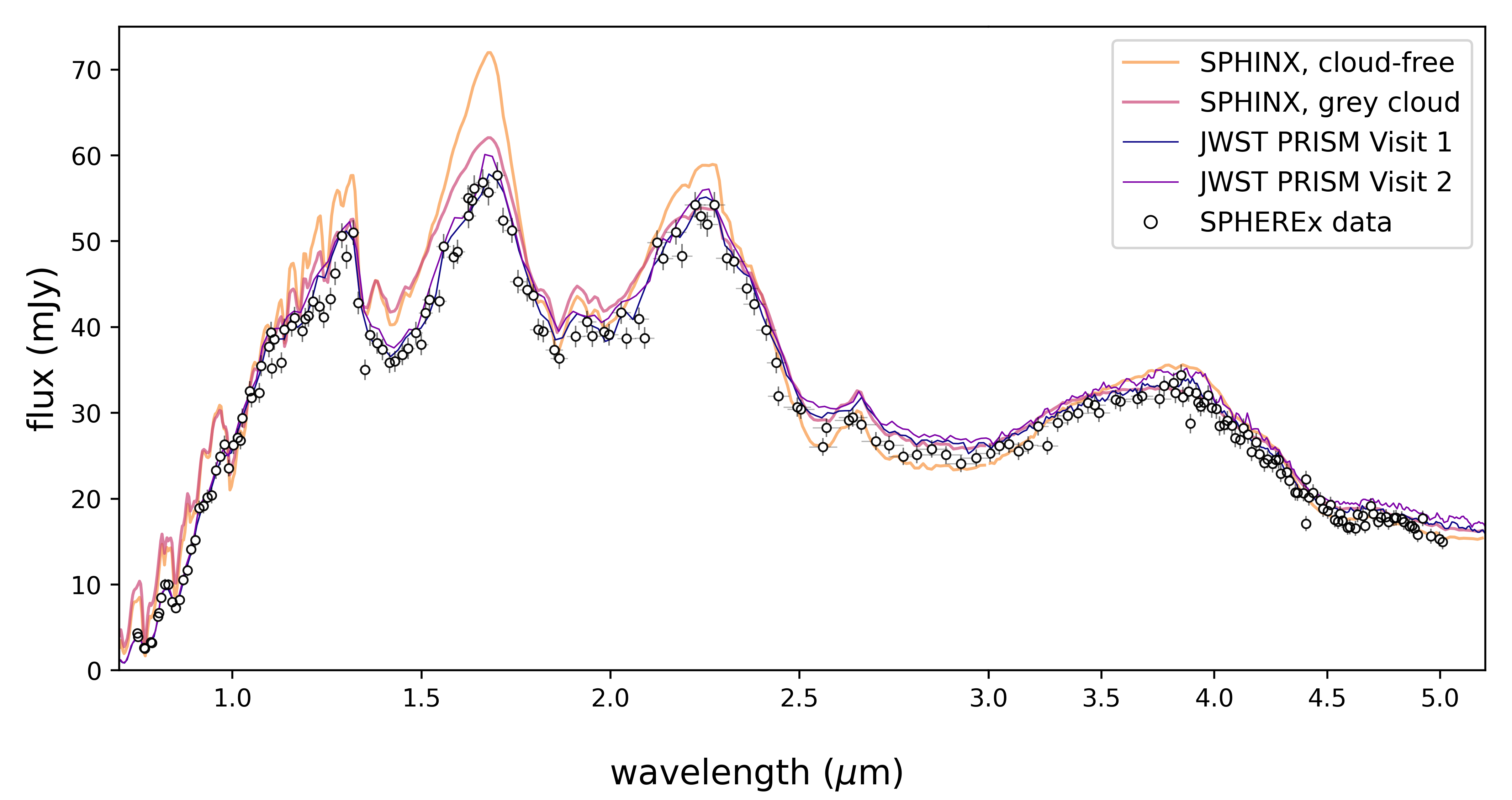}
    \caption{Publicly available SPHEREx spectrum (black circles) of the notable UCD TRAPPIST-1, overlaid with two JWST/NIRSpec-PRISM observations from 2022 (blue and purple lines). The SPHEREx data agree with both JWST spectra, but show systematic tension with the 2600 K cloudless (orange) and cloudy (pink) SPHINX models. These data demonstrate the potential for SPHEREx to inform models of UCDs.}
    \label{fig:trappist1}
\end{figure}

The SPHEREx mission recently completed the first spectroscopic map of the whole sky from space, and these data are now available to the community. The wide, $0.75-5.0\mu m$ wavelength coverage of SPHEREx encapsulates $>90\%$ of the radiation emitted by stars spanning the entire Main Sequence, making it well-suited to measure largely model-independent bolometric fluxes with a common calibration source. Its low-resolution spectra consist of 102 channels distributed over six detectors, capturing the broad molecular absorption features shaping M dwarf spectra. When paired with astrometry from Gaia, SPHEREx spectra can yield refined temperature and radius estimates for hundreds of thousands of nearby M dwarfs.

\begin{figure}
\centering
\includegraphics[width=\linewidth]{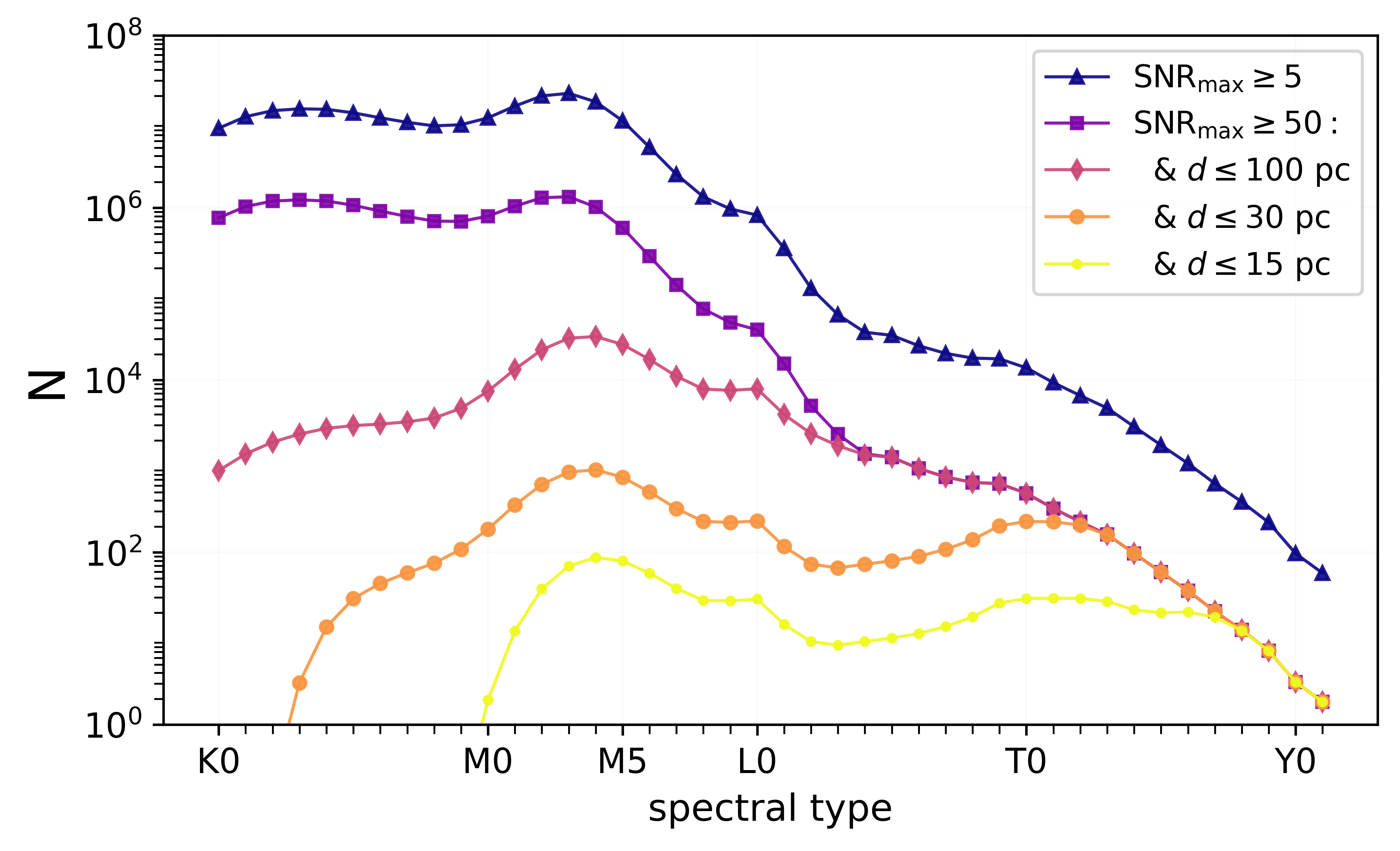}
\caption{The estimated number of stars characterized by SPHEREx as a function of spectral type. The yellow line represents the number of stars of a given spectral type expected to achieve an SNR $\geq$ 5 in any of its 102 channels. The yellow-green line shows the number of stars expected to reach an SNR $\geq$ 50. The lines below show the same SNR threshold, but with 100-, 30-, and 15-pc distance cuts applied. SPHEREx will measure the spectra of $\sim10^6$ M5-M9 dwarfs with SNR $\geq$ 50 and tens of thousands of brown dwarfs, thousands of which are near enough for detailed characterization.}
\label{fig:spherex_yields}
\end{figure}

Given the relatively bright saturation limit of SPHEREx ($\sim 11.7-9.7$ AB mag), most nearby M dwarfs are expected to partially or completely saturate, resulting in poor and often unusable flux measurements. In order to forecast the potential for SPHEREx to characterize POKEMON targets, we use their 2MASS and WISE magnitudes to estimate their AB magnitudes in each of the six SPHEREx detector bands, and we compare these estimates to the published saturation limits of SPHEREx\footnote[4]{Table 16 of the SPHEREx Explanatory Supplement}. We estimate the spectral types of the POKEMON targets by spline-interpolating the $G$-$Ks$ relation in Table 1 of \cite{HenryJao2024ARA&A..62..593H} and using the $G-K_s$ color of each POKEMON target. We find that only 58 of the 520 stars are expected not to saturate in any of the SPHEREx channels, the majority of which are later than M4.5. 138 stars are expected to saturate in all channels, including Bands 5 and 6, which have more lenient saturation limits (Figure \ref{fig:saturation}).

However, SPHEREx can be a powerful tool for the POKEMON targets that do not saturate. The notable POKEMON-DLC target and UCD TRAPPIST-1 -- an exoplanet-hosting M8 at 12 pc -- has now been observed by SPHEREx in two epochs separated by six months. We extracted the SPHEREx spectrum of TRAPPIST-1 using its proper-motion propagated position and the public IRSA Spectrophotometry Tool\footnote[5]{\url{https://irsa.ipac.caltech.edu/applications/spherex/tool-spectrophotometry}}. The spectrum is Nyquist-complete in wavelength, meaning the spectral points are a half-bandwidth apart. We added a 4$\%$ flux uncertainty floor in quadrature with the pipeline errors to account for the visit-to-visit scatter typically seen in the data. The data closely match two independent high-precision observations with JWST/NIRSpec-PRISM (Z. Rustamkulov \& S. Peacock et al., submitted), with qualitative agreement at the uncertainty level of the data (Figure \ref{fig:trappist1}). We compare these data to two state-of-the-art SPHINX models \citep{Iyer2023, Iyer2025}, with and without clouds. These two models give $\chi^2_\nu$ = 8.6 and 10.0, respectively. We find that the addition of clouds somewhat improves the goodness of fit, especially in the 1-2.5 micron range, but the 5$\%$ model deviations inside and between molecular features reflect long-standing modeling challenges that confound our understanding of the atmospheres and interiors of UCDs. However, SPHEREx will shed new light on these objects, promising a statistical understanding of the mysterious red dwarfs at the foot of the hydrogen burning limit.

SPHEREx will also shed new light on a multitude of low-mass stars and brown dwarfs at tens or even hundreds of parsecs. Using the measured 5$\sigma$ spectral sensitivity floor of SPHEREx (19.6-17.5 AB mag\footnote[6]{\href{https://irsa.ipac.caltech.edu/data/SPHEREx/docs/SPHEREx_Expsupp_QR.pdf}{SPHEREx Explanatory Supplement}}) and stellar atmosphere models \citep{Kurucz2005, Allard2014, Iyer2023}, we calculated the SNR of each spectral subtype as a function of distance. Combining these with our inferred space densities, and assuming stars are uniformly distributed in the thin disk with a vertical scale height of 300 pc, we estimate that SPHEREx will measure the spectra of $\sim10^6$ M5-M9 dwarfs with SNR $\geq$50 (we consider a spectrum measured if any one of the 102 SPHEREx channels reaches the SNR requirement). A few thousand of these M dwarfs are ``nearby'' (within a couple dozen pc), putting them well within the reach of precision astrometry, exoplanet searches, stellar multiplicity investigations, rotation rate estimations, and flare studies, without the confounding effects of interstellar extinction (Figure \ref{fig:spherex_yields}). Excitingly, we predict that SPHEREx will also discover and spectroscopically characterize tens of thousands of brown dwarfs, thousands of which are near enough for detailed characterization.

Unfortunately, the use of SPHEREx for studying stellar multiplicity is more limited. The spatial resolution of SPHEREx is $6.2\arcsec$ \citep{Condon2024ApOpt..63.3453C}, so any binary with a projected separation less than this value will be blended in the SPHEREx spectrophotometry. Unblending these sources is non-trivial; for instance, an M5+M5 binary could masquerade as a star that is $\sqrt{2}$ larger in radius. Furthermore, the shape and magnitude of the flux contribution of a blended brown dwarf component would be dwarfed by the $\sim5\%$ spectral deviations from M dwarf models. This result emphasizes that although SPHEREx will provide a myriad of tools for the characterization of low-mass stars, high-resolution imaging remains necessary for the identification of close-in companions.

\section{Conclusions and Future Work} \label{sec:conclusions}

Here we present diffraction-limited speckle observations of \pokemondlc{} nearby M dwarfs, in which we detect four likely bound companions. These observations increase the number of ultracool (later than M6.5) dwarfs in the 15-pc POKEMON catalog by a factor of 1.6. In total, the catalog now consists of \pokemonnum{} stars, of which \multiplenum{} are multiples. There are a total of \companionnum{} companions in these \multiplenum{} multi-star systems.

We also investigated stellar companions known to the literature to present an updated projected separation distribution that has a peak at \projseppeak{}. We also present updated stellar multiplicity and companion rates for M dwarfs of \multiplicityrate{} and \companionrate{}, respectively.

Additionally, we discuss Gaia as an identifier of nearby, ultracool dwarfs, and present expectations for SPHEREx. While many nearby M dwarfs will saturate, SPHEREx will be a powerful tool for the characterization of UCDs. SPHEREx will measure the spectra of $\sim10^6$ M5-M9 dwarfs and tens of thousands of brown dwarfs with SNR $\geq$50. Thousands of these are near enough for detailed characterization. However, SPHEREx will struggle to disentangle multi-star systems due to its moderate ($6.2\arcsec$) spatial resolution. As such, high-resolution will continue to be needed to disentangle stellar pairs.

We are in the process of using spectral energy distribution fitting to homogenously characterize the POKEMON catalog. In forthcoming publications we will present a new tool for rapid characterization of low-mass stars, as well as a catalog of effective temperatures, bolometric luminosities, and radii -- and thus masses -- for a volume-limited sample of nearby M dwarfs. These measurements will provide updated mass ratio and orbital period distributions for M dwarfs, and allow for the stellar multiplicity rate of M dwarfs within 15 pc to be calculated as a function of mass through M9 for the first time.

\begin{acknowledgments}
\indent We are so grateful to the multitude of folks who have contributed to the POKEMON survey, and in particular to Elliott Horch, Zach Hartman, Joe Llama, Mike Lund, Brian Mason, Brian Skiff, Jen Winters, and Frederick Hahne. We also thank the referee for their thoughtful and thorough review of this work.

This research was supported by NSF Grant No.~AST-1616084 and NASA Grant 18-2XRP18\_2-0007.

These results are also based on observations from Kitt Peak National Observatory, the NSF's National Optical-Infrared Astronomy Research Laboratory (NOIRLab Prop. ID: 2024B-520128; PI: C. Clark), which is operated by the Association of Universities for Research in Astronomy (AURA) under a cooperative agreement with the National Science Foundation. The authors are honored to be permitted to conduct astronomical research on Iolkam Du’ag (Kitt Peak), a mountain with particular significance to the Tohono O’odham. Data presented herein were obtained at the WIYN Observatory from telescope time allocated to NN-EXPLORE through the scientific partnership of the National Aeronautics and Space Administration, the National Science Foundation, and the NSF's National Optical-Infrared Astronomy Research Laboratory. Observations in the paper made use of the NN-EXPLORE Exoplanet and Stellar Speckle Imager (NESSI). NESSI was funded by the NASA Exoplanet Exploration Program and the NASA Ames Research Center. NESSI was built at the Ames Research Center by Steve B. Howell, Nic Scott, Elliott P. Horch, and Emmett Quigley.

Some of the observations in the paper made use of the High-Resolution Imaging instruments ‘Alopeke and Zorro. ‘Alopeke and Zorro were funded by the NASA Exoplanet Exploration Program and built at the NASA Ames Research Center by Steve B. Howell, Nic Scott, Elliott P. Horch, and Emmett Quigley. ‘Alopeke and Zorro were mounted on the Gemini North and South telescope of the international Gemini Observatory, a program of NSF’s NOIRLab, which is managed by the Association of Universities for Research in Astronomy (AURA) under a cooperative agreement with the National Science Foundation. on behalf of the Gemini partnership: the National Science Foundation (United States), National Research Council (Canada), Agencia Nacional de Investigación y Desarrollo (Chile), Ministerio de Ciencia, Tecnología e Innovación (Argentina), Ministério da Ciência, Tecnologia, Inovações e Comunicações (Brazil), and Korea Astronomy and Space Science Institute (Republic of Korea).

All the JWST data used in this paper can be found in MAST: \dataset[10.17909/qk94-yj96]{http://dx.doi.org/10.17909/qk94-yj96}.

This work presents results from the European Space Agency (ESA) space mission
. Gaia data are being processed by the Gaia Data Processing and Analysis Consortium (DPAC). Funding for the DPAC is provided by national institutions, in particular the institutions participating in the Gaia MultiLateral Agreement (MLA). The Gaia mission website is \url{https://www.cosmos.esa.int/gaia}. The Gaia archive website is \url{https://archives.esac.esa.int/gaia}.

This work has used data products from the Two Micron All Sky Survey \citep{https://doi.org/10.26131/irsa2}
, which is a joint project of the University of Massachusetts and the Infrared Processing and Analysis Center at the California Institute of Technology, funded by NASA and NSF.

This publication makes use of data products from the Wide-field Infrared Survey Explorer \citep[WISE;][]{https://doi.org/10.26131/irsa1}, which is a joint project of the University of California, Los Angeles, and the Jet Propulsion Laboratory/California Institute of Technology, funded by the National Aeronautics and Space Administration.

This publication makes use of data products from the Spectro-Photometer for the History of the Universe, Epoch of Reionization and Ices Explorer \citep[SPHEREx;][]{https://doi.org/10.26131/irsa652}, which is a joint project of the Jet Propulsion Laboratory and the California Institute of Technology, and is funded by the National Aeronautics and Space Administration.

Information was collected from several additional large database efforts: the Simbad database and the VizieR catalogue access tool, operated at CDS, Strasbourg, France; the Astrophysics Data System operated by the Smithsonian Astrophysical Observatory under NASA Cooperative Agreement 80NSSC25M7105; the Washington Double Star Catalog maintained at the US Naval Observatory; and the fourth US Naval Observatory CCD Astrograph Catalog
.
\end{acknowledgments}

%
\facilities{WIYN(NESSI), Gemini South(Zorro)}

\software{Astropy \citep{Astropy2013}, IPython \citep{IPython2007}, Matplotlib \citep{Matplotlib2007}, NumPy \citep{NumPy2020}, Pandas \citep{Pandas2010}}



\appendix 

\section{Updated Tables} \label{appendix}

In Appendix \ref{appendix}, we update relevant tables from \citet{Clark2024AJ....167...56C} and \citet{Clark2024AJ....167..174C} to include the POKEMON-DLC targets. Table \ref{table:all_targets} lists all \pokemonnum{} targets in the catalog and whether they have detected or known stellar companions. Detected companions are listed in Table \ref{table:all_detected}, and known companions are listed in Table \ref{table:all_known}. In addition to the newly added targets, Table \ref{table:all_detected} now includes one corrected WDS identifier, and Table 6 now includes corrections to several typographical errors in the projected separation values, which had been propagated into the projected separation distribution. We hope that this volume-limited census of nearby M-dwarf multiples will be of use to both the stellar astrophysics and exoplanet communities interested in the Solar Neighborhood.

\begin{deluxetable}{cccccc}
\tablecaption{Targets in the volume-limited, 15-pc POKEMON catalog of M-dwarf primaries
\label{table:all_targets}}
\tablehead{\colhead{2MASS ID or Name} & \colhead{Gaia DR3 ID} & \colhead{TIC ID} & \colhead{Detected companion(s)?} & \colhead{Known companion(s)?} & \colhead{POKEMON-DLC?}}
\startdata
J00064325-0732147 & 2441630500517079808 & 176287658 & N & N & N \\
J00085512+4918561 & 393621524910343296 & 201784071 & N & N & N \\
J00113182+5908400 & 423027104407576576 & 452882423 & N & N & N \\
J00152799-1608008 & 2368293487261055488 & 12862728 & Y & Y & N \\
J00153905+4735220 & 392350008432819328 & 440065990 & N & N & N \\
J00154919+1333218 & 2768048564768256512 & 52005579 & N & N & N \\
J00182256+4401222 & 385334230892516480 & 440109725 & N & Y & N \\
J00202922+3305081 & 2863419584886542080 & 57984826 & N & N & N \\
J00242463-0158201 & 2541756977144595712 & 244167275 & N & N & N \\
J00244419-2708242 & 2322561156529549440 & 340703996 & Y & Y & N \\
\enddata
\tablecomments{Table \ref{table:all_targets} is published in its entirety in the machine-readable format. A portion is shown here for guidance regarding its form and content.}
\end{deluxetable}

\begin{deluxetable}{ccccccccccc}
\tablecaption{Detected companions to the targets in the volume-limited, 15-pc POKEMON catalog of M-dwarf primaries
\label{table:all_detected}}
\tablehead{\colhead{2MASS ID or name} & \colhead{WDS ID} & \colhead{Epoch} & \colhead{$\lambda$} & \colhead{$\Delta\lambda$} & \colhead{$\theta_1$} & \colhead{$\rho_1$} & \colhead{$\Delta m_1$} & \colhead{$\theta_2$} & \colhead{$\rho_2$} & \colhead{$\Delta m_2$} \\ 
\colhead{} & \colhead{} & \colhead{(2000+)} & \colhead{(nm)} & \colhead{(nm)} &\colhead{($^{\circ}$)} & \colhead{($\arcsec$)} & \colhead{(mag)} & \colhead{($^{\circ}$)} & \colhead{($\arcsec$)} & \colhead{(mag)}}
\startdata
J00074264+6022543 & 00077+6022 & 18.5887 & 562 & 44 & 107.6 & 0.9750 & 0.95 &  &  & \\
 &  & 18.5887 & 832 & 40 & 107.9 & 0.9774 & 0.76 &  &  & \\
J00085391+2050252 & 00089+2050 & 18.5853 & 832 & 40 & 232.4 & 0.1329 & 0.64 &  &  & \\
J00152799-1608008 & 00155-1608 & 17.8025 & 692 & 40 & 131.6 & 0.1944 & 2.46 &  &  & \\
 &  & 17.8025 & 880 & 50 & 131.1 & 0.1910 & 1.64 &  &  & \\
 &  & 18.6571 & 832 & 40 & 4.8 & 0.2146 & 1.58 &  &  & \\
J00244419-2708242 & 00247-2653 & 18.6571 & 832 & 40 & 51.7 & 0.9122 & $<2.41$ &  &  & \\
J00322970+6714080 & 00321+6715 & 18.5887 &  562 & 44 & 354.7 & 0.4695 & 3.94 & 185.4 & 3.5795 & $<2.78$ \\
 &  & 18.5887 & 832 & 40 & 355.3 & 0.4806 & 3.05 & 185.6 & 3.5909 & $<2.46$ \\
 &  & 19.0643 & 832 & 40 & 3.8 & 0.4941 & 2.44 &  &  &  \\
\enddata
\tablecomments{Table \ref{table:all_detected} is published in its entirety in the machine-readable format. A portion is shown here for guidance regarding its form and content.}
\tablecomments{Astrometric and photometric uncertainties are described in Sections 3.2 and 3.3, respectively, of \citet{Clark2024AJ....167...56C}.}
\end{deluxetable}

\begin{deluxetable}{cccccccccccccc}
\tablecaption{Known stellar companions to the targets in the volume-limited, 15-pc POKEMON catalog of M-dwarf primaries}
\label{table:all_known}
\tablehead{\colhead{2MASS ID or name} & \colhead{Component} & \colhead{Epoch} & \colhead{$\theta$} & \colhead{Error} & \colhead{$\rho$} & \colhead{Error} & \colhead{$r$} & \colhead{Technique} & \colhead{Reference} & \colhead{$\Delta m$} & \colhead{Error} & \colhead{Filter} & \colhead{Reference} \\ 
\colhead{} & \colhead{} & \colhead{(yr)} & \colhead{($^{\circ}$)} & \colhead{($^{\circ}$)} & \colhead{($\arcsec$)} & \colhead{($\arcsec$)} & \colhead{(au)} & \colhead{} & \colhead{} & \colhead{(mag)} & \colhead{(mag)} & \colhead{} & \colhead{}}
\startdata
J00152799-1608008 & A-B & 1985 & 130 &  & 0.22 &  & 1.0945 & micdet & 31 & 0.4 &  & V & 31 \\
J00182256+4401222 & A-B & 1907 & 57 &  & 38.83 &  & 138.33 & micdet & 11 & 3.02 &  & V & 95 \\
J00244419-2708242 & A-BaBb & 1993 & 353 & 0.4 & 1.07 & 0.4 & 8.2743 & spkdet & 57 & 1.2 &  & K & 57 \\
 & Ba-Bb & 1993 & 330 & 0.4 & 0.267 & 0.4 & 2.0647 & spkdet & 57 & 0.35 &  & K & 57 \\
J00275592+2219328 & A-B & 2004 & 13 & 2 & 0.125 & 0.01 & 1.77 & AO det & 23 & 0.26 & 0.05 & K & 23 \\
J00322970+6714080 & AaAb-B & 1923 & 99 &  & 2.26 &  & 22.511 & phodet & 93 & 2.5 &  & V & 21 \\
 & Aa-Ab & 1989 & 43 & 3 & 0.451 & 0.02 & 4.4922 & spkdet & 70 & 1.94 & 0.06 & K & 70 \\
LTT 10301 & A-B & 1960 & 315 &  & 1 &  & 14.986 & phodet & 61 & 0.3 &  & B & 61 \\
J00582789-2751251 & A-B &  &  &  &  &  & 0.2381 & SB1 & 24 &  &  &  &  \\
J01023895+6220422 & A-B & 1999 & 76 &  & 293.05 &  & 2889.4 & visdet & 81 & 2.35 & 0.03 & K & 81 \\
\enddata
\tablecomments{Table \ref{table:all_known} is published in its entirety in the machine-readable format. A portion is shown here for guidance regarding its form and content.}
\tablecomments{The codes for the techniques and instruments used to detect and resolve systems are: AO det—detection via adaptive optics; astdet—detection via astrometric perturbation, companion often not detected directly; astorb—orbit from astrometric measurements; CCDdet—detection via CCD or other two-dimensional electronic imaging; lkydet—detection via lucky imaging; micdet-detection via a micrometry technique; occdet-detection via occulation; phodet—detection via a photographic technique; radvel—detection via radial velocity, but no SB type indicated; SB (1, 2, 3)—spectroscopic multiple, either single-lined, double-lined, or triple-lined; spcdet-detection via space-based technique; spkdet—detection via speckle interferometry; visdet—detection via wide-field CCD or other two-dimensional electronic imaging.}
\tablerefs{(1) \citet{Aitken1899AN....150..113A}; (2) \citet{Balega1999AAS..140..287B}; (3) \citet{Baroch2018AA...619A..32B}; (4) \citet{Benedict2000AJ....120.1106B}; (5) \citet{Beuzit2004AA...425..997B}; (6) \citet{Blazit1987AAS...71...57B}; (7) \citet{Bowler2015ApJS..216....7B}; (8) \citet{Burnham1891AN....127..289B}; (9) \citet{Burnham1891AN....127..369B}; (10) \citet{Burnham1894PLicO...2....1B}; (11) \citet{Burnham1913mpms.book.....B}; (12) \citet{Clark2023RNAAS...7..206C}; (13) \citet{Cortes-Contreras2017AA...597A..47C}; (14) \citet{Dahn1976PUSNO..24c...1D}; (15) \citet{Delfosse1999AA...344..897D}; (16) \citet{Delfosse1999AA...350L..39D}; (17) \citet{DuquennoyMayor1988AA...200..135D}; (18) \citet{El-Badry2021MNRAS.506.2269E}; (19) \citet{ESA1997ESASP1200.....E}; (20) \citet{Espin1920MNRAS..80..329E}; (21) \citet{EspinMilburn1926MNRAS..86..131E}; (22) \citet{Forrest1988ApJ...330L.119F}; (23) \citet{Forveille2005AA...435L...5F}; (24)
\citet{Fouque2018MNRAS.475.1960F}; (25) \citet{Franz1998AJ....116.1432F}; (26) \citet{Gaia2023AA...674A...1G}; (27) \citet{Giclas1971lpms.book.....G}; (28) \citet{Gili2022AN....34324008G}; (29) \citet{GizisReid1996AJ....111..365G}; (30)   \citet{Hartkopf1994AJ....108.2299H}; (31) \citet{Heintz1987ApJS...65..161H}; (32) \citet{Heintz1993AJ....105.1188H}; (33) \citet{Henry1997AJ....114..388H}; (34) \citet{Henry1999ApJ...512..864H}; (35) \citet{Henry2018AJ....155..265H}; (36) \citet{HerbigMoorhead1965ApJ...141..649H}; (37) \citet{Hertzsprung1909AN....180...39H}; (38) \citet{Horch2004AJ....127.1727H}; (39) \citet{Horch2010AJ....139..205H}; (40) \citet{Horch2012AJ....143...10H}; (41) \citet{Hough1899AN....149...65H}; (42) \citet{Hussey1904LicOB...2..180H}; (43) 
\citet{Janson2012ApJ...754...44J}; (44) \citet{Janson2014ApJ...789..102J}; (45) \citet{Jeffers2018AA...614A..76J}; (46) \citet{Jodar2013MNRAS.429..859J}; (47) 
\citet{Joy1942PASP...54...33J};(48) \citet{Joy1947ApJ...105...96J}; (49) \citet{Kuiper1934PASP...46..360K}; (50) \citet{Kuiper1936ApJ....84..478K}; (51) \citet{Kuiper1936ApJ....84R.359K}; (52) \citet{Kuiper1942ApJ....96..315K}; (53) \citet{Kuiper1943ApJ....97..275K}; (54) \citet{Lamman2020AJ....159..139L}; (55) \citet{Lau1911AN....189..197L}; (56) \citet{Leinert1986AA...164L..29L}; (57) \citet{Leinert1994AA...291L..47L}; (58) \citet{Luyten1941POMin...3....1L}; (59) \citet{Luyten1969PMMin..21....1L}; (60) \citet{Luyten1972PMMin..29....1L}; (61) \citet{Luyten1977PMMin..50....1L}; (62) \citet{Luyten1979nltt.book.....L}; (63) \citet{Luyten1997yCat.1130....0L}; (64) \citet{Luyten1998yCat.1087....0L}; (65) \citet{Malo2014ApJ...788...81M}; (66) \citet{Martin2000ApJ...529L..37M}; (67) \citet{Mason2018AJ....155..215M}; (68) \citet{McCarthyHenry1987ApJ...319L..93M}; (69) \citet{McCarthy1988ApJ...333..943M}; (70) \citet{McCarthy1991AJ....101..214M}; (71) \citet{Montagnier2006AA...460L..19M}; (72) \citet{OsvaldsOsvalds1959AJ.....64..265O};(73) \citet{ReidGizis1997AJ....113.2246R}; (74)
\citet{ReinersBasri2010ApJ...710..924R}; (75) \citet{Reuyl1941PASP...53..119R}; (76) \citet{Richichi1996AA...309..163R}; (77) 
\citet{Riedel2010AJ....140..897R}; (78) \citet{Rossiter1955POMic..11....1R}; (79) \citet{Salama2021AJ....162..102S}; (80) \citet{Shkolnik2010ApJ...716.1522S}; (81) \citet{Skrutskie2006AJ....131.1163S}; (82) 
\citet{Sperauskas2019AA...626A..31S}; (83) \citet{Struve1837AN.....14..249S}; (84) \citet{Tokovinin2010AJ....139..743T}; (85) \citet{TomkinPettersen1986AJ.....92.1424T}; (86) \citet{vanBiesbroeck1961AJ.....66..528V}; (87) \citet{vanBiesbroeck1974ApJS...28..413V}; (88) \citet{vandeKamp1936PASP...48..313V}; (89) \citet{vandenBos1937CiUO...98..362V}; (90) \citet{vandenBos1950CiUO..109Q.371V}; (91) \citet{vandenBos1951CiUO..111...13V}; (92) \citet{Vrijmoet2022AJ....163..178V}; (93) \citet{Vyssotsky1927PA.....35..213V}; (94) \citet{Ward-Duong2015MNRAS.449.2618W}; (95) \citet{Wendell1913AnHar..69...99W}; (96) \citet{Wilson1954AJ.....59..132W}; (97) \citet{Winters2017AJ....153...14W}; (98) \citet{Wirtanen1941PASP...53..340W}; (99) \citet{Worley1962AJ.....67..403W}.}
\end{deluxetable}

\clearpage

\bibliography{sample701}{}
\bibliographystyle{aasjournalv7}



\end{document}